\documentclass[12pt]{article}

\usepackage[utf8]{inputenc}
\usepackage{cite}

\usepackage{graphicx}
\usepackage{subcaption}
\usepackage{caption}

\usepackage{amsmath,latexsym,amssymb}
\usepackage[colorlinks=true,citecolor=magenta,urlcolor=blue]{hyperref}

\usepackage{hyperref}

\usepackage{geometry}
\newgeometry{vmargin={32mm}, hmargin={32mm,32mm}}

\begin{document}

\vspace*{-14mm}

\begin{center}
  {\Large\bf{Tsung-Dao Lee has died, \vspace*{4mm} \\
   long live parity symmetry breaking!}}
  \vspace*{5mm} \\

{\small Wolfgang Bietenholz \\
Instituto de Ciencias Nucleares \\
Universidad Nacional Aut\'onoma de M\'exico (UNAM) \\
Apartado Postal 70-543, 04510 Ciudad de M\'exico, M\'exico} \vspace*{2mm} \\

\end{center}

\noindent
On August 4 this year, Tsung-Dao Lee, a renowned theoretical physicist of
Chinese origin, passed away at the age of 97. His most famous discovery
dates back to 1956, when -- together with Chen-Ning Yang -- he postulated
that parity symmetry might be broken by the weak interaction. They
suggested experimental tests of this revolutionary idea,
which were conducted within one year. The results confirmed
the conjecture by Lee and Yang, thus changing a core paradigm
of physics.

\section{Is there a difference between left and right?}

Imagine that you could talk on the phone to an extraterrestrial.
The translation machine works well, each one narrates about his
civilization, until a tricky question arises: you want to
explain what we call ``left'' and ``right'', but how do you do it?
It is not a video call, and he doesn't know our anatomy, so saying
``the left is where we have the heart'' or
``the right-handed sugar molecules are the ones that our body digests''
or something similar doesn't work.
Is it even possible to explain this? Can you propose an
experiment whose outcome could answer this question?

Until 1956, it was a core paradigm in physics that the laws of Nature
are {\em parity invariant}; in the framework of Quantum Mechanics, this
was pointed out by Eugene Wigner \cite{Wigner}.
A parity transformation ($P$-transformation)
means that the three spatial coordinates change sign,
$P: \ \vec r \to - \vec r$, but we can instead
simply imagine that we are observing the world in a mirror. Then,
left and right are exchanged, but apparently the physical
processes persist. For example, if we watch a billiard game in a
mirror, it seems that nothing deviates from the natural laws. In fact,
gravitational and electromagnetic forces, and even the
strong interaction of Quantum Chromodynamics, remain invariant.
If $P$-invariance were universally valid, there would be no way to
explain to the extraterrestrial what we mean by ``left'' and ``right''. 

In 1956, however, this paradigm was challenged by Tsung-Dao Lee and
Chen-Ning Yang, two young Chinese in New York, aged 29 and 33, who
wondered: does parity symmetry also hold for the {\em weak interaction},
which, for example, causes the radioactive decay? It was standard to
assume that this was obvious and that it had already been tested, but
a critical review of the known experiments with the weak
interaction showed that none of them had actually
verified parity symmetry \cite{LeeYangP}. On the contrary,
Lee and Yang found an indication that it was broken.

\begin{figure}[h!]
\centering
\begin{subfigure}{0.35\textwidth}
	\includegraphics[scale=.8]{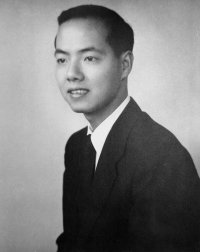}
\end{subfigure}	
\begin{subfigure}{0.5\textwidth}
  \includegraphics[scale=0.5]{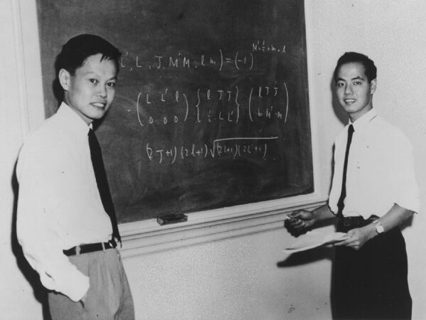}
\end{subfigure}	
\caption*{\hspace*{1cm}
Left: Tsung-Dao Lee, physics Nobel Prize laureate at the age of 30.\\
\hspace*{1cm}
Right: Chen-Ning Yang (left) and Tsung-Dao Lee (right) in Princeton\\
\hspace*{1cm} (credit to Alan Richards).}
\end{figure}

At that time, there was extensive discussion about the
so-called ``$\theta^{+}$--$\tau^{+}$ puzzle'':
these two unstable particles had been observed in cosmic ray
experiments since 1947. They are mesons with
positive electric charge, one was assumed to decay into
two pions and the other into three pions (pions, $\pi$, are the
lightest mesons),
$$
\theta^+ \to \pi^+ + \pi^0 \quad , \qquad
\tau^+ \to \pi^+ + \pi^+ + \pi^-
\quad {\rm or} \quad  \pi^+ + \pi^0 + \pi^0 \ . 
$$
Pions have an intrinsic parity of $-1$, meaning that under a parity
transformation, their fields change sign. This implies that the state of
two pions has positive parity, $(-1)^2 = 1$, whereas the state of three
pions has negative parity, $(-1)^3 = -1$ (at zero orbital
angular momentum). Therefore, it was assumed that $\theta^+$
had the intrinsic parity $+1$, and that it was $-1$ for $\tau^+$.

But the puzzle was that the measurements indicated (within their
precision) the same mass (494~MeV) and mean lifetime
($1.2 \cdot 10^{-8}$~s) for $\theta^{+}$ and $\tau^{+}$,
which seemed like a strange coincidence.
Lee and Yang correctly hypothesized that it was the {\em same}
meson \cite{LeeYangP}, which we now call a {\em kaon}, $K^+$. It
has parity $-1$, but it can decay into two {\em or} three pions (among
other decay channels) because its decay involves the weak interaction
(therefore the lifetime is relatively long), which does {\em not}
conserve parity. To be explicit: its purely hadronic decay modes have
the branching ratios $\pi^+ + \pi^0$: 20.7\%; $\pi^+ + \pi^+ + \pi^-$:
5.6\% and $\pi^+ + \pi^0 + \pi^0$: 1.8\%.

In their paper, Lee and Yang suggested a multitude of experiments to
directly test parity violation in the weak interaction \cite{LeeYangP}.
Their first and simplest proposal refers to the $\beta$-decay of
a neutron into a proton, an electron, and an anti-neutrino,
$$
n \to p^+ + e^- + \bar \nu_e \ .
$$
The neutron can be part of an unstable nucleus; Lee and Yang proposed
$^{60}_{27}$Co.
The spin of the nucleus has the direction of its magnetic moment, which
can be aligned by applying a magnetic field. Spin is another intrinsic
degree of freedom of particles, which behaves like angular momentum,
$\vec{L} = \vec{r} \times \vec{p}$. Under parity transformation,
the position vector $\vec{r}$ and the momentum $\vec{p}$ change
sign, so $\vec{L}$ is invariant, as is spin. However, the momenta
of the electron and the anti-neutrino do change sign.

This means that, if the $\beta$-decay is parity invariant, the
electron flux -- which is measurable -- in the direction of the spin
and in the opposite direction should be equally intense. Already
in 1956, this experiment was carried out under the direction of
Chinese-American physicist Chien-Shiung Wu (she cancelled a planned
journey to Europe and Asia in order to work on it as soon as possible).
There were practical challenges: to achieve significant polarization
of the nuclei, a strong magnetic field was required, along with a
very low temperature of 0.003~K.

\begin{figure}[h!]
\centering
\includegraphics[scale=0.4]{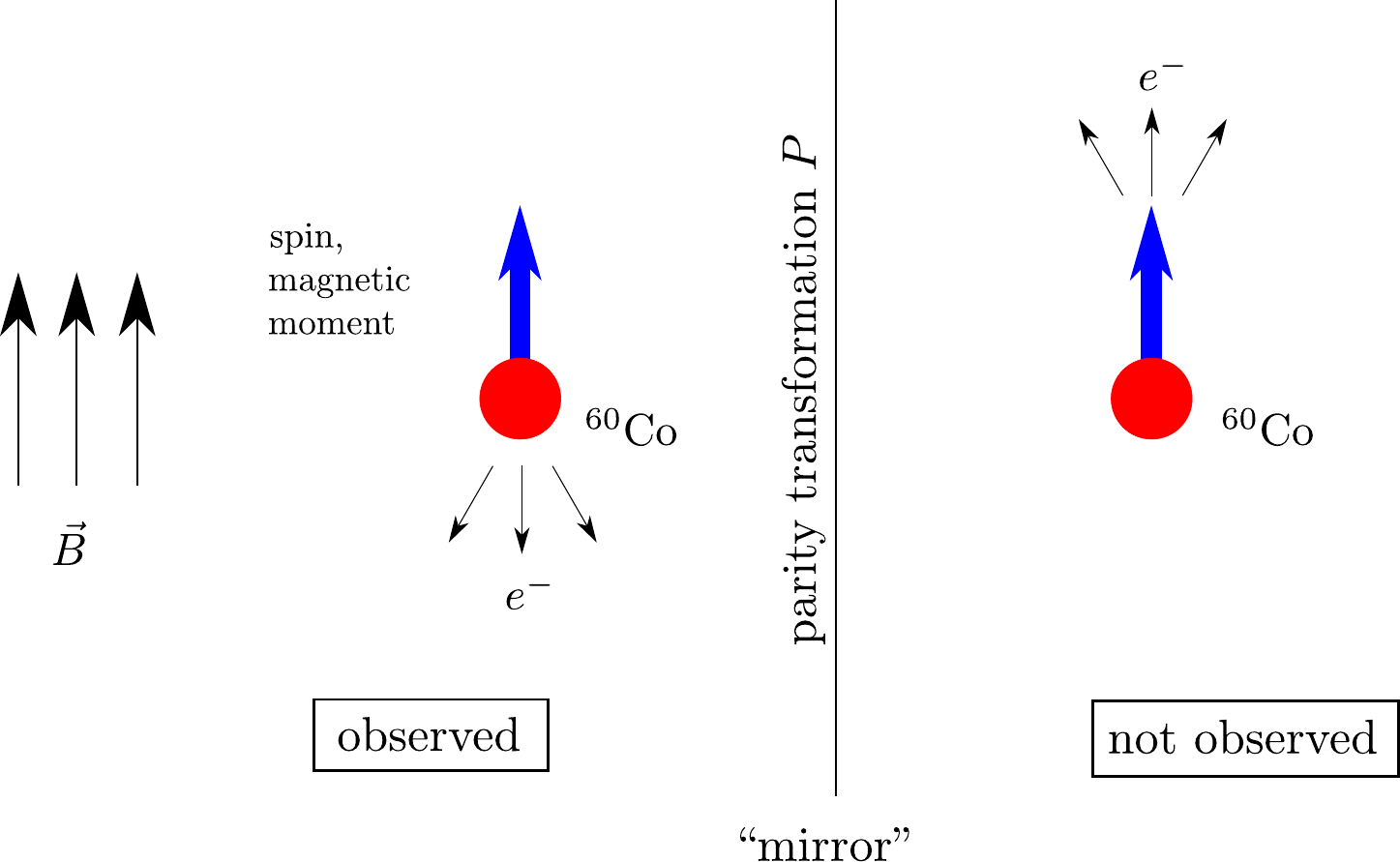}
\caption*{The concept of the Wu experiment: under parity
  transformation, the electron flux flips to the other side of
  the polarized $^{60}_{27}$Co nuclei. So if $P$-invariance holds, this
  flux must be equally intense in both directions. The experiment
  demonstrated that this is {\em not} the case \cite{Wu}.}
\vspace*{-2mm}
\end{figure}

The result for a sample of cobalt nuclei, with about 60\% polarization,
was clear and submitted for publication in January 1957: the electron
flux in the direction opposite to the nuclear spin is stronger, demonstrating
that the weak interaction breaks parity symmetry \cite{Wu}. Somewhat
more detailed descriptions are given in the Appendix and in
Ref.\ \cite{neutrinorev}.

If the extraterrestrial also performs this experiment, he will
be able to distinguish between clockwise and counterclockwise
rotation, and finally understand what we mean by ``left'' and
``right''. This came as a great surprise in 1957; at first,
prominent physicists like Wolfgang Pauli did not believe it,
but it was substantiated by further experiments \cite{Telegdi}.
Still in 1957, Lee and Yang received the Nobel Prize
for this discovery, the same year when Albert Camus won the
Nobel Prize in literature. Lee, at the age of 30, was the youngest
Nobel Prize winner in physics since World War II. Lee and Yang
were the first Nobel laureates from China. Unfortunately,
Madame Wu was not included, which many people consider unjust.
In 1978, she was finally awarded the (inaugural) Wolf Prize in physics
for this achievement.

\begin{figure}[h!]
\centering
\includegraphics[scale=0.27]{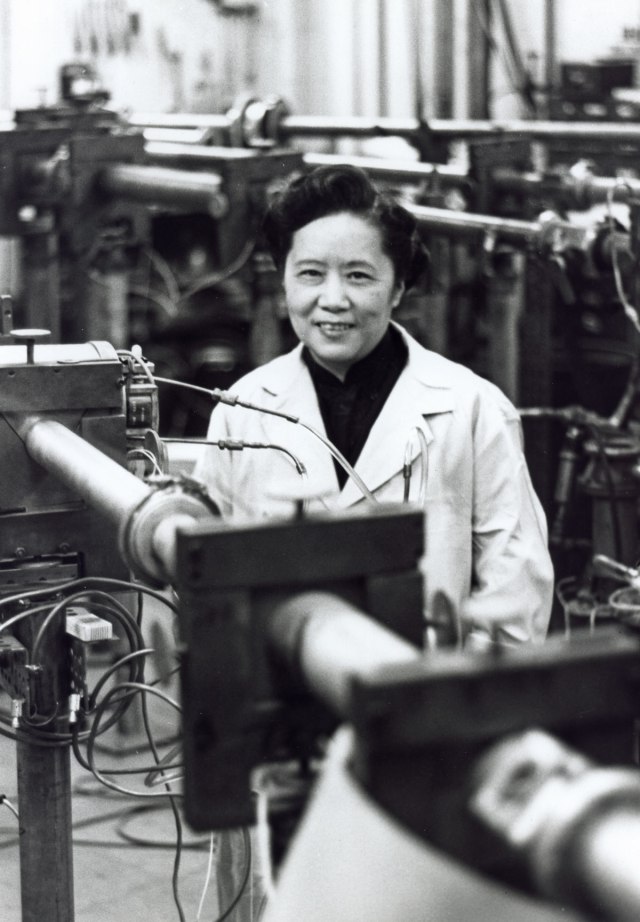}
\hspace*{1.5cm}
\includegraphics[scale=0.335]{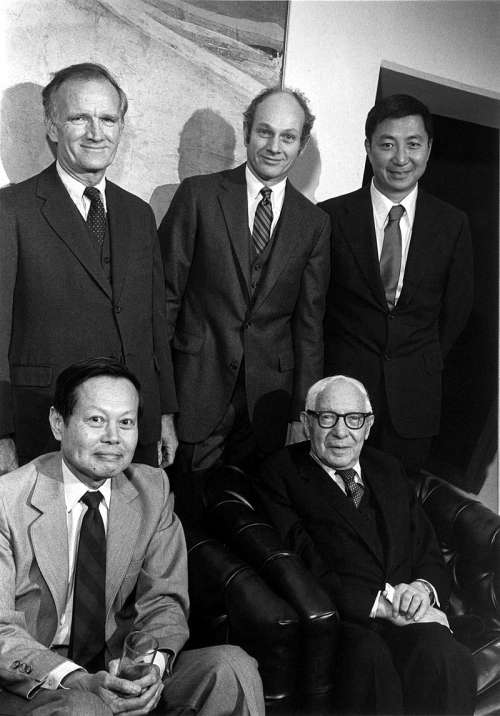}
\caption*{Left: Chien-Shiung Wu in her laboratory.
  Right: (from left to right, standing): Val Fitch, James
  Cronin (they discovered the $CP$-violation \cite{CPviolation}),
  and Samuel Ting; (seated): Chen-Ning Yang and Isidor Rabi.
  They are all Nobel Prize winners.}
\vspace*{-2mm}
\end{figure}

We usually assume matter to dominate all regions of the Universe.
However, if our extraterrestrial friend were made of anti-matter,
he might misinterpret this experiment.
Wu's experiment demonstrated not only the violation of
$P$-invariance, but also of {\em $C$-invariance} \cite{Okun}, which
exchanges matter and anti-matter. In fact, a decaying anti-cobalt
nucleus emits a positron {\em in} the direction of its nuclear spin.
The experiment is indistinguishable, however, under
the combination $CP$, so {\em $CP$-invariance} became the new
paradigm \cite{Landau}. However, just a few years later, in 1964,
it was refuted as well, this time with the decay of neutral
kaons \cite{CPviolation}.

What is still considered valid today is {\em $CPT$-invariance},
which additionally reverses the direction of time (like a movie
played backwards). There are good reasons for this property to hold:
in a quantum field theory -- the successful formalism of particle
physics -- that is local and Lorentz invariant, $CPT$-invariance
must hold \cite{CPT}. The clearest proof was provided by Res Jost
by means of analytic continuation, in a groundbreaking paper, which
he wrote in German and published in a journal that no longer exists
\cite{Jost}.

\newpage

\section{Who was Tsung-Dao Lee?}

Lee -- known as T.-D.\ Lee -- was born in 1926 in Shanghai, where he
grew up and first studied chemical engineering and then physics.
Due to the Japanese invasion, he left Shanghai and
continued his studies in Kunming.
After the war, he received a fellowship from the Chinese
government under Chiang Kai-shek, which was granted to very few
exceptionally gifted students to study in the United States.
The goal was to promote the development of nuclear weapons in China
upon their return \cite{Wang}. Thus, Lee arrived 1946 in Chicago,
where he completed his Ph.D.\ thesis about White Dwarfs under the
supervision of Enrico Fermi in 1950, one year after the Chinese Revolution.

Yang, who was born in Hefei in 1922, also arrived in Chicago in 1946,
where he obtained his Ph.D.\ in 1948 under the direction of
Edward Teller, then he worked as an assistant of Fermi. In 1952,
Lee and Yang published together two papers that still play an important
role in Statistical Mechanics \cite{LeeYang0}. They deal with the
density of zeros of a partition function, depending on an external
source, as the volume increases, which serves as an indicator of
phase transitions.

In 1956, the year when they conjectured the breaking of parity invariance,
Lee already became a full professor at Colombia University. In the same
year, Lee and Yang also introduced a more abstract, discrete
symmetry, the {\em $G$-parity} \cite{Gparity}. Formally,
a $G$-transformation is obtained by combining a $C$-transformation
with an ``isospin rotation'' (from a modern point of view, this is
a ``rotation'' between the lightest quark flavors $u$ and $d$).
It amounts to a sign flip of the pion fields, without
performing a spatial inversion (the latter differs from a
$P$-transformation).
The strong interaction, limited to these two quark flavors,
is $G$-invariant, therefore it cannot change the number of pions
from even to odd, or vice versa.

Until 1962, Lee and Yang published together more than 30 papers
that received great recognition. Apparently, the collaboration
was lively: colleagues of the time heard the two often
shouting at each other, alternating between Chinese and English
\cite{Nature}. Later, there were disagreements related to the
merits of these successful works, and ever since they have
gone separate ways.

Both continued to feel connected to their home country. After
U.S.\ President Nixon visited China in 1972, bilateral relations
improved, and it also became possible for other US citizens
to visit China. Lee and Yang immediately took this opportunity
to undertake lecture tours (Yang already in 1971), where they
were celebrated as heroes. Lee was received by Premier Zhou Enlai
in 1972, and in 1974 also by Chairman Mao Zedong, who wanted to
discuss with him the question what symmetries mean
from a philosophical perspective \cite{LeeBooklet}.
Mao regretted not having studied more science. With Zhou's
support, Lee advocated for the resumption of natural science
education during the Cultural Revolution.

Lee continued to promote China's investment in basic research.
Later he was in contact with Deng Xiaoping and acted as an
(unofficial) scientific advisor. In particular, he promoted the
construction of the {\it Beijing Electron Positron Collider} (BEPC),
which went into operation in 1989.

In contrast, Yang recommended that China
focus on applied research, as well as social and environmental issues.
In 2016, this led him to reject the planned {\it Circular Electron
Positron Collider} (CEPC, a ``Higgs factory'' of 100~km circumference
and center-of-mass energies up to 240~GeV). Not surprisingly, this
provoked objections from the active generation of Chinese physicists,
who expect a beneficial boost in science and technology, if the
construction of the CEPC will be approved \cite{YangCEPC}.

\begin{figure}[h!]
\centering
	\includegraphics[scale=0.16]{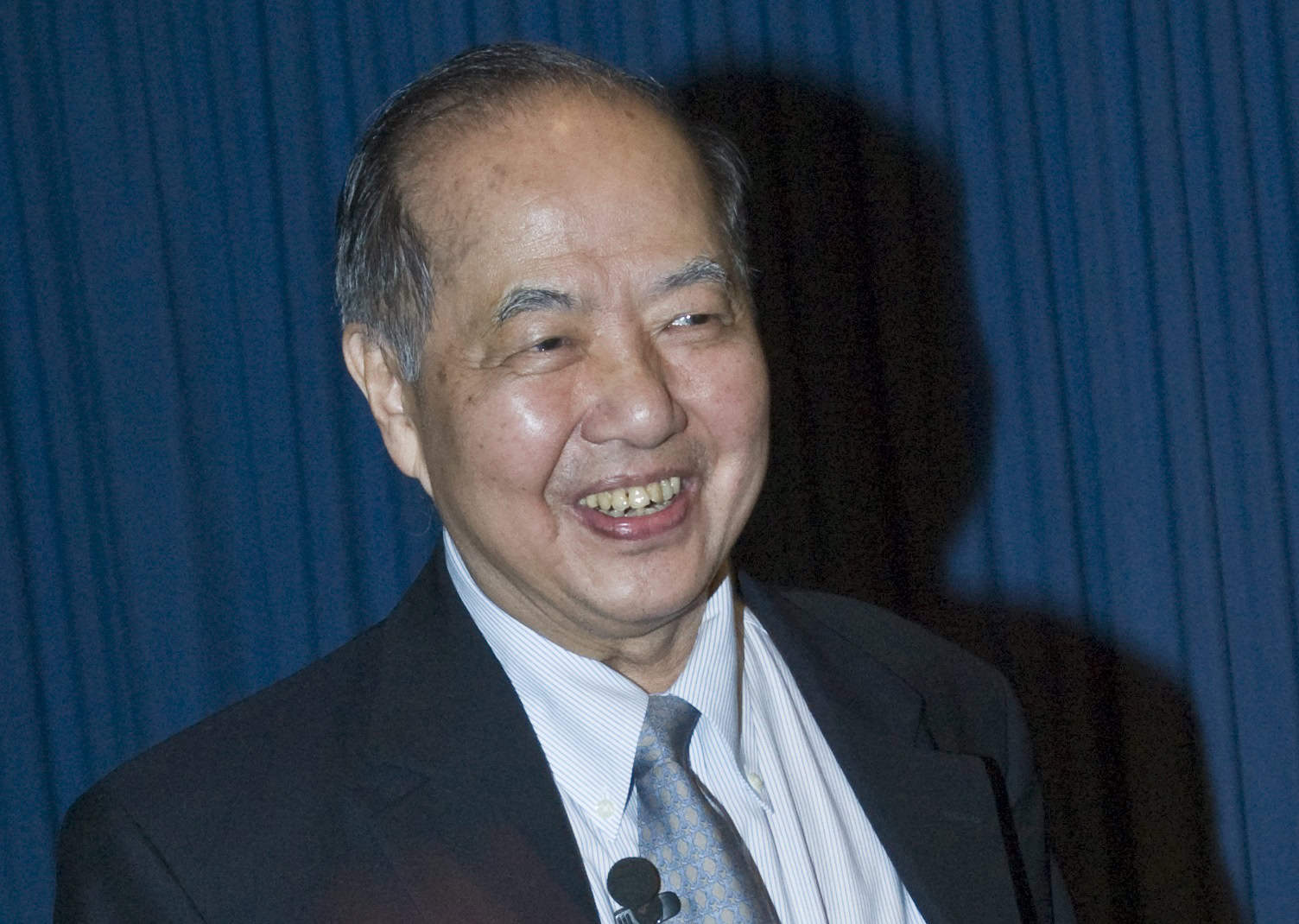}
\caption*{Tsung-Dao Lee visiting CERN.} 
\end{figure}

Yang taught at Stony Brook University in New York from 1965 to
1999. He is also famous for non-Abelian gauge theory
(Yang-Mills theory) and the Yang-Baxter equation, a consistency
condition for scattering processes. Yang is 102 years old and
he now lives in Beijing.\\

I myself have encountered Lee on two occasions. First, when I
was a {\it summer student} at CERN and Lee gave a 
colloquium on parity violation, which I heard again years later
as a postdoc at MIT. He also published this well-designed and
pedagogical lecture as a booklet \cite{LeeBooklet}, in which a
variant of intergalactic communication about {\it left} and {\it right}
is discussed already. The lecture starts with the picture of two cars
that are mirror images of each other, so one would be allowed in most
countries, while the other in countries like the UK or Japan.
If parity invariance were universally valid, both cars would
be equally fast, but if the weak interaction plays a role in
the engine, this does not need to be true anymore.
Additionally, Lee wrote an extensive textbook on particle
physics \cite{LeeBook}.

At MIT, Lee also gave a seminar for the lattice group, proposing
a new lattice field theory formulation, which, however, barely
attracted interest. On the other hand, there was no doubt about
his personal authority. When we met and shook hands, he simply
said, with a friendly smile, ``How do you do?'', well aware that
he did not need to introduce himself. When a colleague in the
seminar commented: ``I guess there is a non-locality in this formulation'',
he responded: ``Don’t try to guess, just follow the talk''.

In the colloquium, someone raised an objection and added: ``There are
many people who think that Lorentz invariance could be broken''. Lee
again gave a decisive response: ``Well, if you want to decide a
scientific question through a popular referendum $\dots$''. Indeed,
speculations about a possible breaking of Lorentz invariance became
popular in the following years, but to this day, there is no evidence
of it.

Despite the humorous formulation of his response, it is an interesting
question how paradigm shifts actually occur in science. There are no
votes or committees for this. A standard reference for the dynamics of
such processes, from a sociological perspective, is a book by Thomas Kuhn
\cite{Kuhn}.
These are rare events and therefore particularly important, like in
1956/7 when the seemingly sacred law of parity invariance in Nature
was overturned.

Another major achievement, where Lee was involved,
is known as the {\em Kinoshita-Lee-Nauenberg Theorem} \cite{KLN}.
It ensures that the infrared divergences cancel in the perturbative
expansion of a class of quantum theories, which includes Quantum
Electrodynamics and (as we know today) even the Standard Model.
In fact, we are usually concerned only with ultraviolet divergences.

Together with Gian Carlo Wick -- a famous theoretical physicist from
Italy, who had been Fermi's assistant in Rome in the 1930s -- Lee
proposed an effective model for unusual states of heavy nuclei
\cite{LeeWick}. They coupled the nucleons to a scalar field,
which could take, in a limited region, a metastable state,
associated with a non-global potential minimum. At strong coupling,
the energy minimum of the system may temporarily favor a kind of
exotic state of the nucleus, with a reduced ``effective'' nucleon mass.
This work had some impact in the conceptual development of the
quark-gluon plasma theory.

As the first director of the RIKEN-Brookhaven National Laboratory,
from 1997 to 2003, Lee supported the funding of a teraflop
supercomputer and, later, a 10-teraflop supercomputer for the
lattice group. He taught at Columbia University from 1953 to 2012,
that is, until the age of 85. \\

Lee was one of the most distinguished theoretical physicists
of the 20th century. In addition to the Nobel Prize, he
received 14 other major awards, and there is a {\it Tsung-Dao Lee
Institute} at Jiao Tong University in Shanghai.
On top of that, he was active in the artistic field: he conceptualized
two sculptures representing ``The Tao of All Matter'' and
Galileo Galilei, which are located in Beijing and Rome,
respectively \cite{Nature}. As a theoretical physicist,
Lee always sought the connection with experiments, following
Fermi's instruction. He expressed this principle in his colloquium
with two rules: {\em ``Without experimentalists, theorists tend to
drift. Without theorists, experimentalists tend to falter''}
\cite{LeeBooklet}. We should remember his advice.

\begin{figure}[h!]
\vspace*{-5mm}
\centering
\includegraphics[scale=0.22]{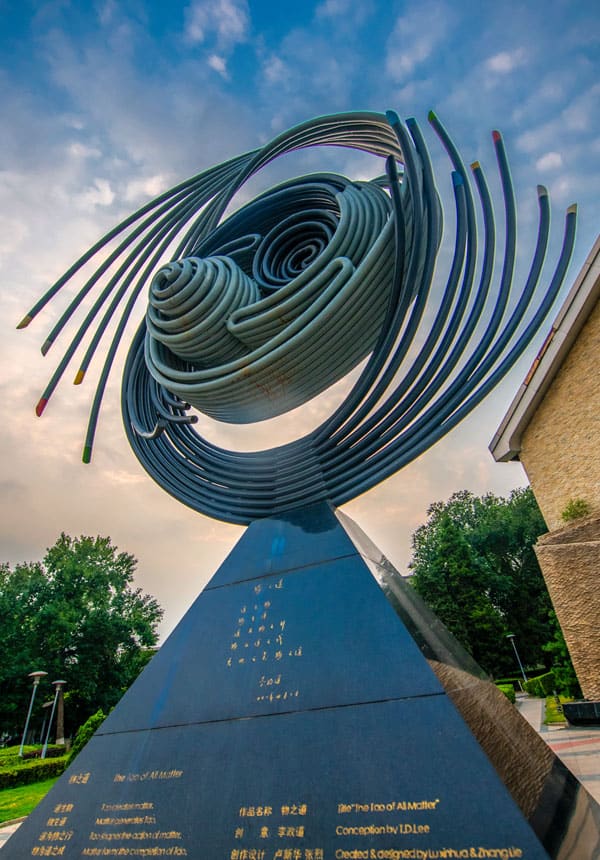}
\caption*{A 5-meter-tall sculpture representing Yin and Yang,
  related to a cyclotron-type accelerator. It is situated in front
  of the Institute of High Energy Physics in Beijing since 2001, and
  it was conceptualized by Tsung-Dao Lee. The left-hand side displays
  the following poem: ``The Tao of All Matter: Tao creates matter,
  matter generates Tao. Tao shapes the action of matter, matter
  forms the completion of Tao. The Art of the universe is the
  Tao of all matter''
  (credit to Institute of High Energy Physics).}
\end{figure}

\newpage
On August 4, 2024, Tsung-Dao Lee passed away in San Francisco at
the age of 97; may he rest in peace.\\

\noindent
{\bf Acknowledgements:} I thank Jos\'{e} Antonio Garc\'{\i}a-Hern\'{a}ndez
for his help with the illustration of the Wu experiment. A shorter version
of this article will be published in Spanish in the {\it Boletín ICNoticias}
of the Instituto de Ciencias Nucleares, UNAM.
I am grateful for the assistance provided by Aline Guevara Villegas,
Carmen Ortega Alfaro and Verenise Sanchez Correa.

\appendix

\section{The Wu experiment from a modern perspective}

Now that we are familiar with the Standard Model of particle physics,
it is rather easy to understand the outcome of the Wu experiment.
First we should mention that the decay transforms the cobalt nucleus
into an excited nickel nucleus,
$$
^{60}_{27}{\rm Co} \quad \to \quad ^{60}_{28}{\rm Ni}^{*} + e^{-} + \bar \nu_e \ ,
$$
thus reducing the nuclear spin from $J=5$ to $4$ (in natural units,
with $\hbar = 1$). The missing nuclear spin is carried away by the
leptons $e^{-}$ and $\bar \nu_e$ (without orbital angular momentum),
which both have spin $1/2$.
Since the electron-anti-neutrino $\bar \nu_e$
is almost massless,
we can safely assume its chirality to coincide with its polarization
(in the nuclear rest system), such that right-handedness (left-handedness)
means that its spin is oriented in (against) its direction of motion.

Thus two scenarios are conceivable:
\begin{itemize}

\item A right-handed anti-neutrino $\bar \nu_{e,R}$ moves in the
  nuclear spin direction, and a left-handed electron $e_L$ in the
  opposite direction.

\item A right-handed electron $e_R$ moves in the nuclear spin direction,
  and a left-handed electron anti-neutrino $\bar \nu_{e,L}$ in the
  opposite direction.
  
\end{itemize}

However, the Standard Model only contains left-handed neutrinos and
right-handed anti-neutrinos, $\nu_L$ and $\bar \nu_R$, in any fermion
generation. This fact alone shows already that the Standard Model
cannot be $P$-invariant, nor $C$-invariant.
Indeed, we have never detected any $\nu_R$ or $\bar \nu_L$.
Even if they exist, they are sterile
Majorana neutrinos, which are not involved in the weak interaction,
therefore they cannot emerge in the $\beta$-decay.
This singles out the first of the two options mentioned above
as the only valid one, which explains the observation in the
Wu experiment \cite{Wu}.

(For completeness, we add that the excited nickel nucleus also emits
two photons, $^{60}_{28}{\rm Ni}^{*} \to \, ^{60}_{28}{\rm Ni} + 2 \gamma$.
That decay is electromagnetic, and therefore $P$-invariant.
The angular $\gamma$-distribution showed to which extent the
nuclei were polarized, and that the $P$-breaking in the
electron-distribution was significant.)

\vspace{-1mm}

\end{document}